\documentclass[a4paper,11pt]{article}
\pdfoutput=1 % if your are submitting a pdflatex (i.e. if you have
\usepackage{epsfig}
\usepackage{graphicx}                  % provides \includegraphics command
\usepackage{ae}
\usepackage{amsmath}
\usepackage{amssymb}
\usepackage{graphics}
\usepackage{dcolumn}
\usepackage{bm}
\usepackage{lineno}
\usepackage{caption}
\usepackage{ragged2e}
\title{\bf Study of charging up effect \\in a triple GEM detector}
\date{}

\begin{document}
		\maketitle
	\flushbottom
\vspace*{-1cm}
\centering
{
\author{S. Chatterjee$^{1*}$, }
\author{A. Sen$^1$,}
\author{S. Roy$^1$,}
\author{K. Nivedita G$^2$,}
\author{A. Paul$^3$,}

\author{S. Das$^1$,}
\author{and S. Biswas$^1$}
}

\vspace*{0.5cm}
$^1${Department of Physics and Centre for Astroparticle Physics and Space Science (CAPSS), Bose Institute, EN-80, Sector-V, Bidhannagar, Kolkata-700091, India}

$^2${{IISER, Maruthamala PO, Vithura, Thiruvananthapuram, Kerala - 695551, India}

$^3${{Department of Physics, University of Calcutta, 92, APC Road, Kolkata - 700009, India}

\let\thefootnote\relax\footnotetext{$^*$Corresponding author. 

\hspace*{0.4cm}E-mail: sayakchatterjee@jcbose.ac.in, sayakchatterjee896@gmail.com }
\vspace*{0.5cm}
\centering{\bf Abstract}
\justify
	The advancement of Micro Pattern Gaseous Detector technology offers us different kinds of detectors with good spatial resolution and high rate capability and the Gas Electron Multiplier~(GEM) detector is one of them. Typically GEM is made up of a thin polyimide foil having a thickness of 50 micrometers with 5 micrometers copper cladding on top and bottom sides. The presence of polyimide changes the gain of the detector under the influence of external radiation and the phenomenon is referred to as the charging up effect. The charging up effect is investigated with a double mask triple GEM detector prototype with Ar/CO$_{2}$ gas mixture in 70/30 ratio under continuous irradiation from a strong Fe$^{55}$ X-ray source. The detailed method of measurements and the test results are presented in this article.

\vspace*{0.25cm}

% \collaboration{\includegraphics[height=17mm]{example-image}\\[6pt]
%   XXX collaboration}
% or

	\section{Introduction}
	\label{sec:intro} 
	
	Gas Electron Multiplier~(GEM) is one of the most advanced detectors of the Micro Pattern Gas Detector~(MPGD) group~\cite{sauli_GEM, sauli_GEM_overview}. GEM is widely used in many High Energy Physics~(HEP) experiments as a tracking device because of its good position resolution due to its micro pattern structure~\cite{compass, TOTEM, ALICE, CMS}. The high rate handling capability of the GEM detector makes it a suitable candidate for the experiments where large particle flux is expected~\cite{CBM_detector}. GEM is made up of a thin Kapton foil of thickness 50~$\mu m$ with 5~$\mu m$ copper cladding on sides of the foil. A large number of holes are etched on the Kapton using the photolithographic technique~\cite{photolithography}.
	
	%%%%%%%%%%%%%%%%%%%%%%%%%%%%%%%%%%%%%%%%%%%%%%%%%    
	\begin{figure}[htbp]
		\centering % \begin{center}/\end{center} takes some additional vertical space
		\vspace*{-2.0cm}
		\includegraphics[scale=0.35]{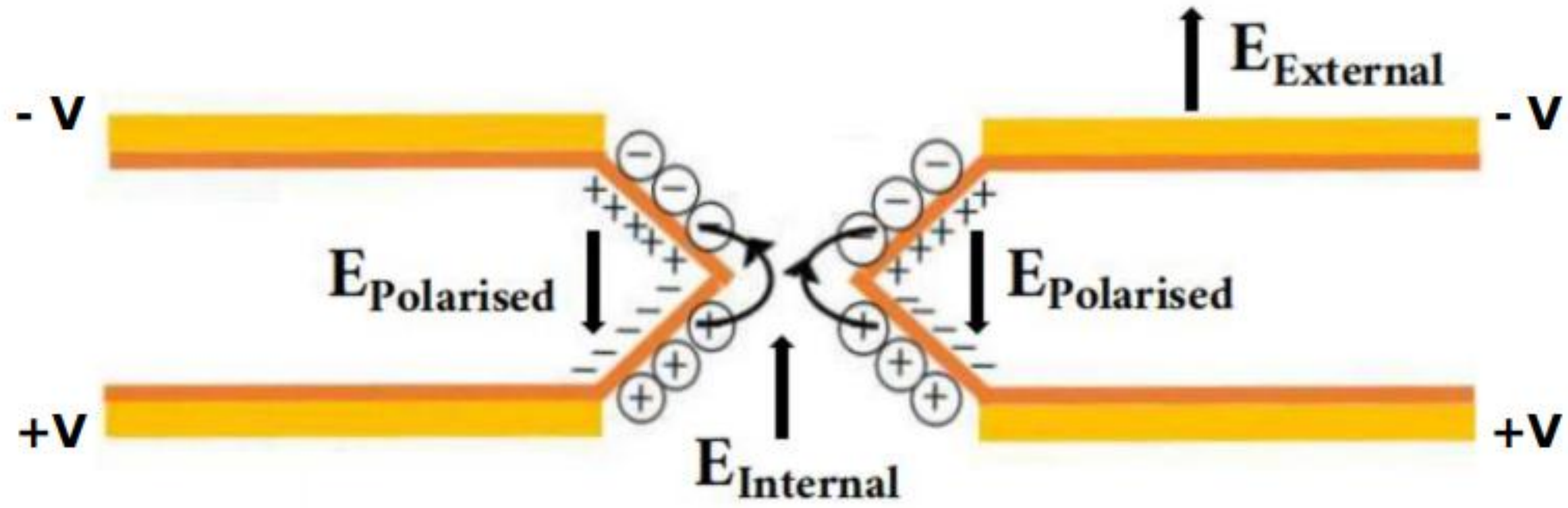}
		\vspace*{-2.0cm}
		\caption{\label{fig:charging_up}Schematic representation of the Charging up effect inside a GEM hole. E$_{Polarised}$ indicates the electric field generated due to the dielectric polarisation. E$_{External}$ indicates the electric field generated due to the external high voltage and E$_{Internal}$ indicates the electric field generated due to the accumulation of the charges on the kapton wall. }
	\end{figure}
	%%%%%%%%%%%%%%%%%%%%%%%%%%%%%%%%%%%%%%%%%%%%%%%%% 
	
	The holes in a standard GEM foil have an outer and inner diameter of 70~$\mu$m and 50~$\mu$m respectively. The distance between the centers of two neighboring holes, that is the pitch, is 140~$\mu$m.
	To create an electric field inside the holes, an external high voltage~(HV) is applied between the copper layers. The holes in the GEM foil act as the multiplication region for the incoming electrons. As shown in Fig.~\ref{fig:charging_up} usually voltage is applied in such a way that the top of the GEM foil is at negative potential compared to the bottom plane and the electrons move downwards. The electrode placed above the top layer of GEM foil is called the drift electrode or drift plane and the gap between the drift plane and top of the GEM foil is called the drift region. An incoming charged particle produces primary electrons mainly in the drift region. These primary electrons are focussed towards the GEM holes by the electric field. The high electric field inside the holes enforces the electrons to multiply by an avalanche. Several GEM layers can be used in cascade mode to attain high gain without increasing the biasing voltage and consequently the discharge probability of the chamber~\cite{bachmann, sbiswas_spark}. 
	
	The presence of the Kapton foil inside the active part of the detector changes its behavior when exposed to external radiation. Due to the high electric field~($\sim kV/cm$) inside the GEM holes, the incoming electrons get sufficient kinetic energy to start an avalanche of further ionization. Due to the dielectric properties of the polyimide~(Kapton), they get polarised by the external electric field. During this multiplication process, the electrons and ions may diffuse to the polyimide surface and due to the polarisation of the polyimide by the external HV, the ions or electrons can be captured on the wall of the Kapton foil. This phenomenon is illustrated in Fig.~\ref{fig:charging_up}. Due to the high resistivity of the Kapton, the charges remain there for a rather long time. As a result of sufficient accumulation of charge on the wall, the electric field configuration inside the hole changes dynamically and this phenomenon is known as the charging up effect. The accumulated charges on the surface of the Kapton foil increase the field inside the holes and as a result, the gain of the chamber increases with time. Many studies have reported that the charging up effect is responsible for a time-dependent change in gain, which asymptotically reaches a constant value~\cite{charging_up_2, charging_up_3, charging_up_4, charging_up_5}. In this article, a systematic investigation of the charging up process with different irradiation rates in a triple GEM detector prototype built using double mask foils operated with Ar/CO$_2$ gas mixture in the 70/30 volume ratio is reported. A strong Fe$^{55}$ source is used to irradiate as well as to record the 5.9 keV X-ray spectra from the chamber. The details of the detector setup are described in Sec.~\ref{setup} and the results are discussed in Sec.~\ref{results}.    
	\section{Detector description and experimental setup}
	\label{setup}
	In this study, a triple GEM detector prototype, consisting of 10~cm~$\times$~10~cm
	double mask GEM foils, obtained from CERN is used. The drift, transfer, and induction gaps of the chamber are kept at 3~mm, 2~mm, and 2~mm respectively (3-2-2-2 configuration).
	
	%%%%%%%%%%%%%%%%%%%%%%%%%%%%%%%%%%%%%%%%%%%%%%%%%    
	\begin{figure}[htbp]
		\centering % \begin{center}/\end{center} takes some additional vertical space
		\vspace*{-.3cm}
		\includegraphics[scale=0.35]{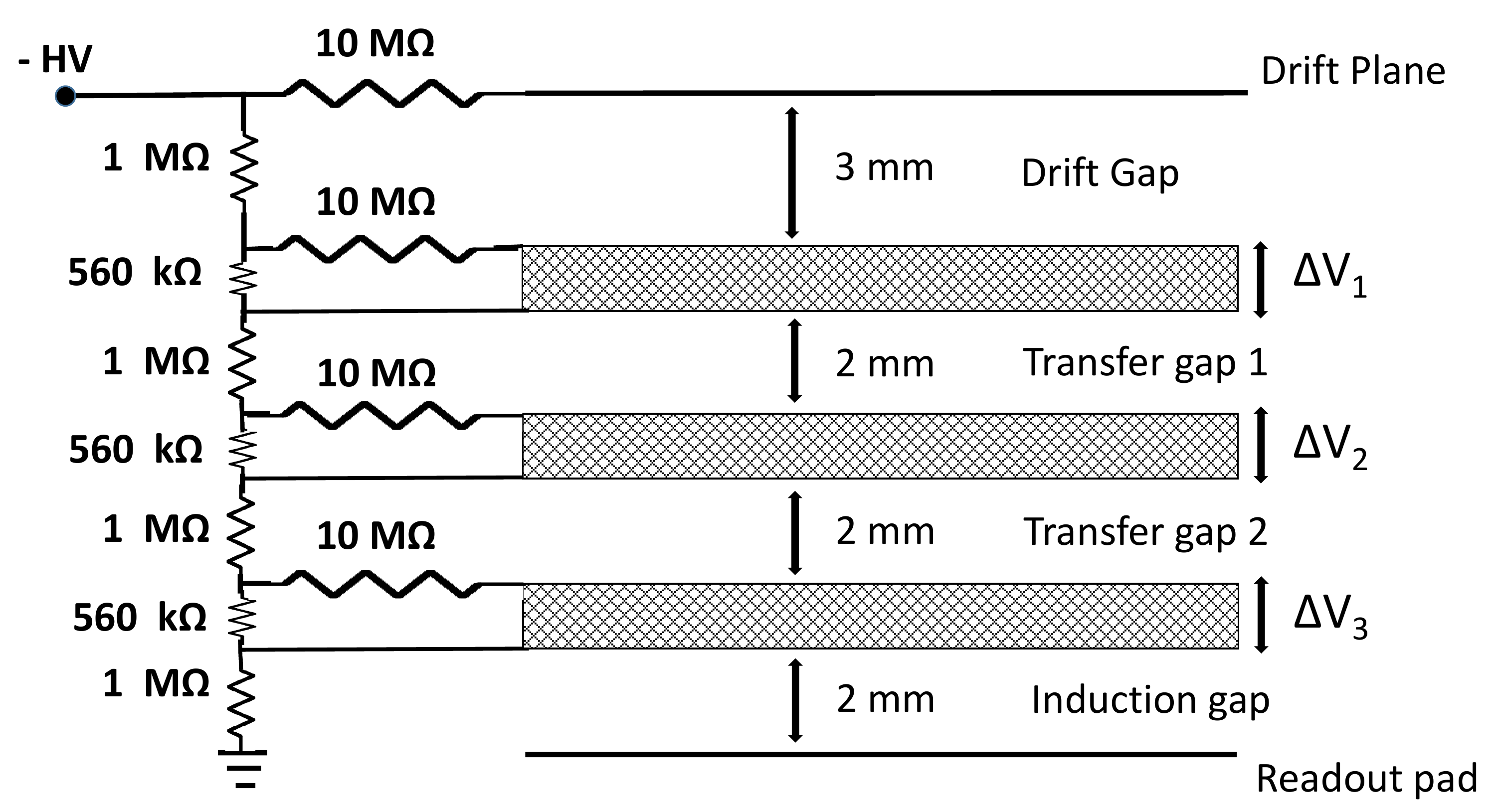}
		\vspace*{-.3cm}
		\caption{\label{fig:detector_setup} Schematic of the HV distribution of the triple GEM chamber. The drift gap, transfer gap and induction gaps are kept at 3~mm, 2~mm, and 2~mm respectively. }
	\end{figure}
	%%%%%%%%%%%%%%%%%%%%%%%%%%%%%%%%%%%%%%%%%%%%%%%%% 
	
	The HV to the drift plane and individual GEM planes are
	applied through a voltage dividing resistor chain. 10~M$\Omega$ protection resistors are applied to the drift plane and top of each GEM foil. A schematic of the resistor chain and different gaps of the chamber is shown in Fig,~\ref{fig:detector_setup}. The readout of the chamber is made up of nine pads of dimension 9~mm~$\times$~9~mm each. The signals in this study are taken from all the pads added by a sum up board and a single input is fed to a charge sensitive preamplifier (VV50-2)~\cite{preamp}. The gain of the preamplifier is 2~mV/fC with a shaping time of 300~ns. A NIM based data acquisition system is used to process the signals from the preamplifier. The output signal from the preamplifier is fed to a linear Fan-in-Fan-out (linear FIFO) module. One analog signal from the linear FIFO is put to a Single Channel Analyser~(SCA) to measure the rate of the incident particle. The SCA is operated in integral mode and the lower level in the SCA is used as the threshold to the signal. The threshold is set at 0.1 V to reject the noise. The discriminated signal from the SCA, which is TTL in nature, is put to a TTL-NIM adapter and the output NIM signal is counted using a NIM scaler. The count rate of the detector in Hz is then calculated. Another output of the linear FIFO is fed to a Multi-Channel Analyser (MCA) to obtain the energy spectra. A schematic representation of the electronics set-up is shown in Fig.~\ref{fig:electronic_setup}.
	
	Pre-mixed Ar/CO${_2}$ gas in a 70/30 volume ratio is used for the whole study.
	A constant gas flow rate of 3.5~l/hr is maintained using a V{\"o}gtlin gas flow meter. Perspex, aluminium and G-10 collimators having different hole diameters are used to irradiate the chamber with different X-ray flux coming from the Fe$^{55}$ source. The ambient temperature, pressure, and relative humidity are monitored continuously using a data logger, built-in house~\cite{data logger}.
	
	%%%%%%%%%%%%%%%%%%%%%%%%%%%%%%%%%%%%%%%%%%%%%%%%%
	\begin{figure}[htbp]
		\centering % \begin{center}/\end{center} takes some additional vertical space
		\includegraphics[scale=0.6]{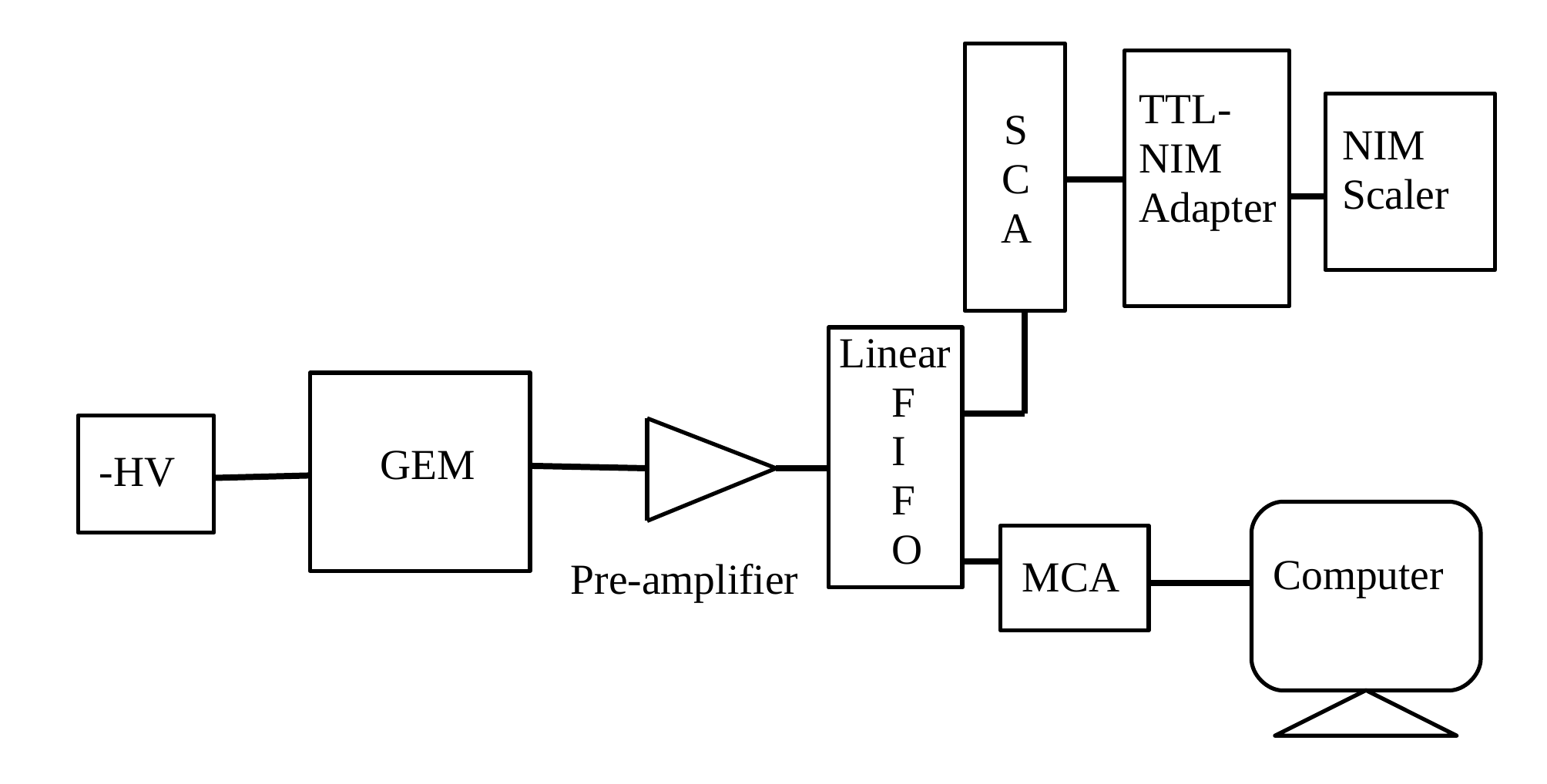}
		\caption{\label{fig:electronic_setup}Schematic representation of the electronics setup}
	\end{figure}
	%%%%%%%%%%%%%%%%%%%%%%%%%%%%%%%%%%%%%%%%%%%%%%%%%
	
	\section{Results}
	\label{results}
	The 5.9~keV peak of the Fe$^{55}$ energy spectrum obtained from the MCA is fitted with a Gaussian distribution to obtain the gain of the chamber. A typical Fe$^{55}$ energy spectrum at -~4.2~kV is shown in Fig.~\ref{fig:fe55_sprctra}. The applied HV of -~4.2~kV corresponds to $\Delta V$ of $\sim$~390~V across each GEM foil and the drift field, transfer field, and induction field of ~2.3~kV/cm,~3.5~kV/cm, and~3.5~kV/cm respectively.  
	
	%%%%%%%%%%%%%%%%%%%%%%%%%%%%%%%%%%%%%%%%%%%%%%%%%    
	\begin{figure}[htbp]
		\centering % \begin{center}/\end{center} takes some additional vertical space
		\includegraphics[scale=0.4]{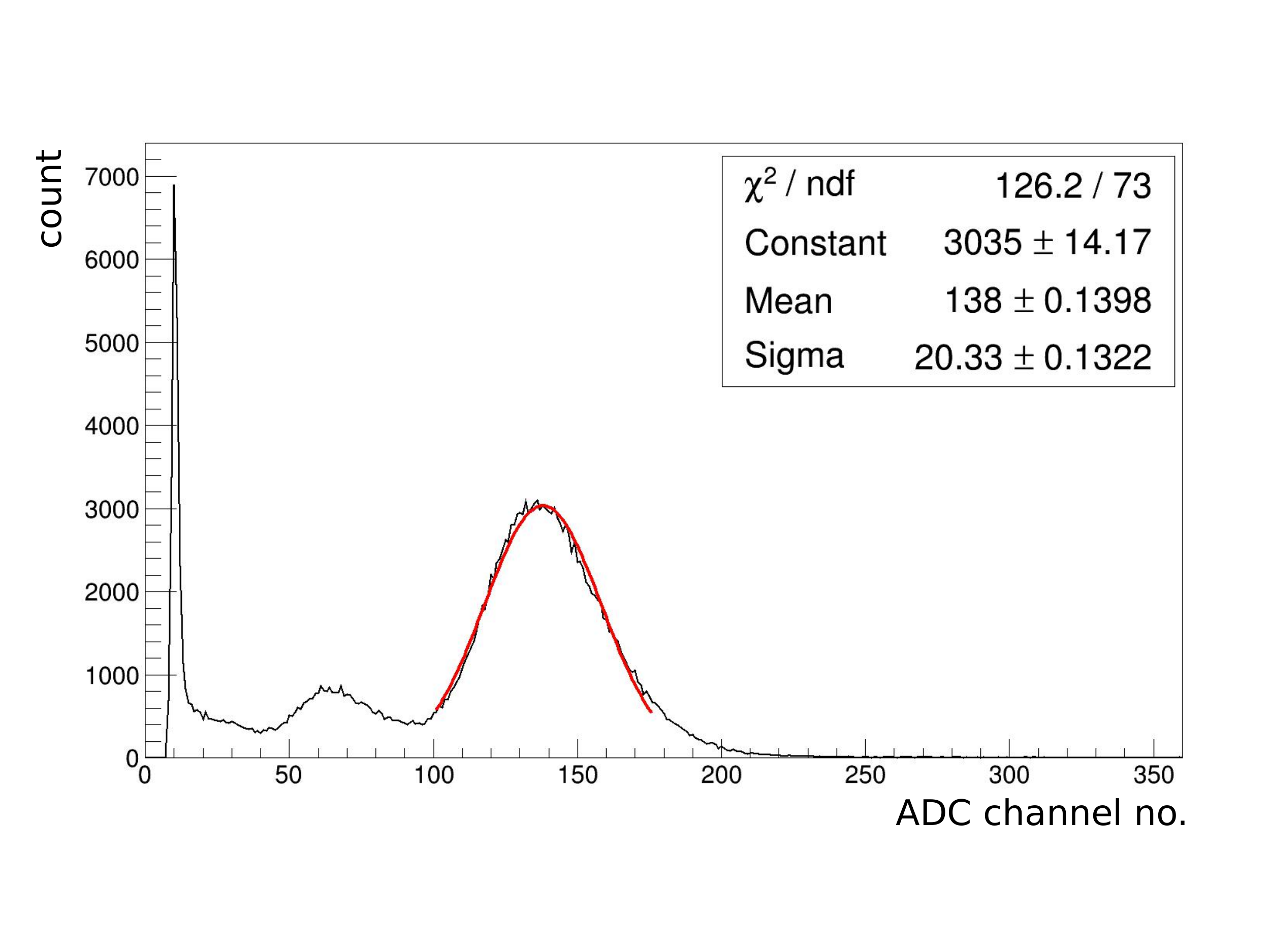}
		\vspace*{-1.0cm}
		\caption{\label{fig:fe55_sprctra}Typical Fe$^{55}$ spectra obtained at a HV of -~4.2~kV and irradiated with 10~kHz X-ray on 50~mm$^{2}$ area . The Main peak is fitted with a Gaussian distribution~(red line) to calculate the gain of the chamber.}
	\end{figure}
	%%%%%%%%%%%%%%%%%%%%%%%%%%%%%%%%%%%%%%%%%%%%%%%%%
	
	The amount of the input charge is calculated by assuming the full energy deposition of the 5.9~keV X-ray in the 3~mm drift gap of the chamber. The number of primary electrons for Ar/CO$_{2}$ in the 70/30 ratio is 212. The ratio of the output charge to the input charge gives the gain of the chamber. The details of the gain calculation and long-term behavior of the chamber were reported earlier~\cite{s. roy,s. chatterjee JOP}.  
	The variation of the gain as a function of time for three different rates of the incoming X-rays, 1~kHz, 10~kHz, and 90~kHz respectively along with the ratio of ambient temperature~(T) to pressure~(p) are shown in the top~(a), middle~(b) and bottom~(c) plot of Fig.~\ref{fig:gain_tp_time}. Using collimators, X-rays of rates 1~kHz, 10~kHz, and 90~kHz are made to fall on 13~mm$^{2}$, 50~mm$^{2}$ and 28~mm$^{2}$ area of the chamber, which implies particle flux of 0.08~kHz/mm$^{2}$, 0.2~kHz/mm$^{2}$ and 3.2~kHz/mm$^{2}$ respectively. All the measurements are carried out at an HV of -~4.2 kV.

	%%%%%%%%%%%%%%%%%%%%%%%%%%%%%%%%%%%%%%%%%%%%%%%%%
	\begin{figure}[htbp]
		\centering % \begin{center}/\end{center} takes some additional vertical space
		\includegraphics[scale=0.4]{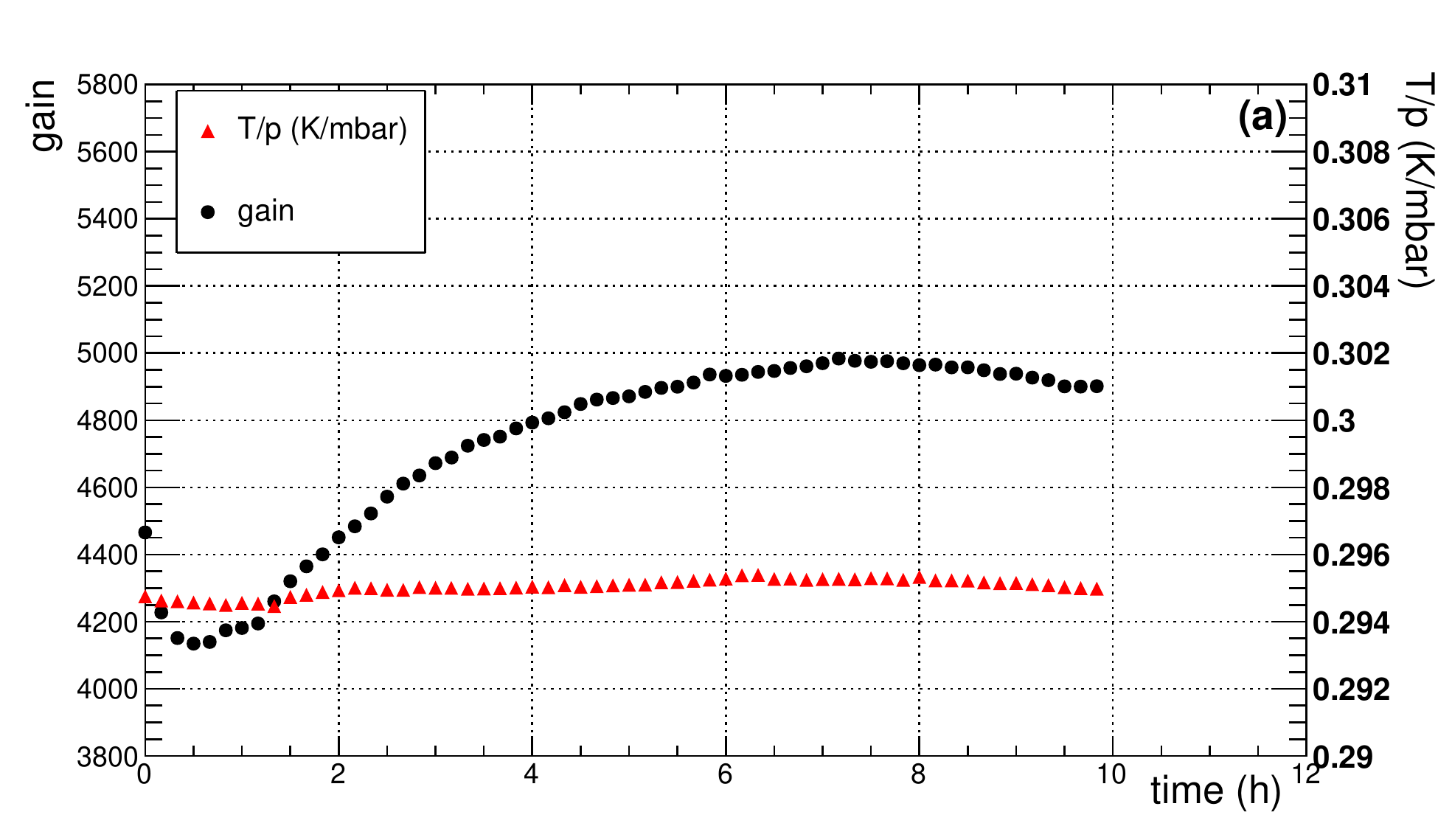}
		\includegraphics[scale=0.4]{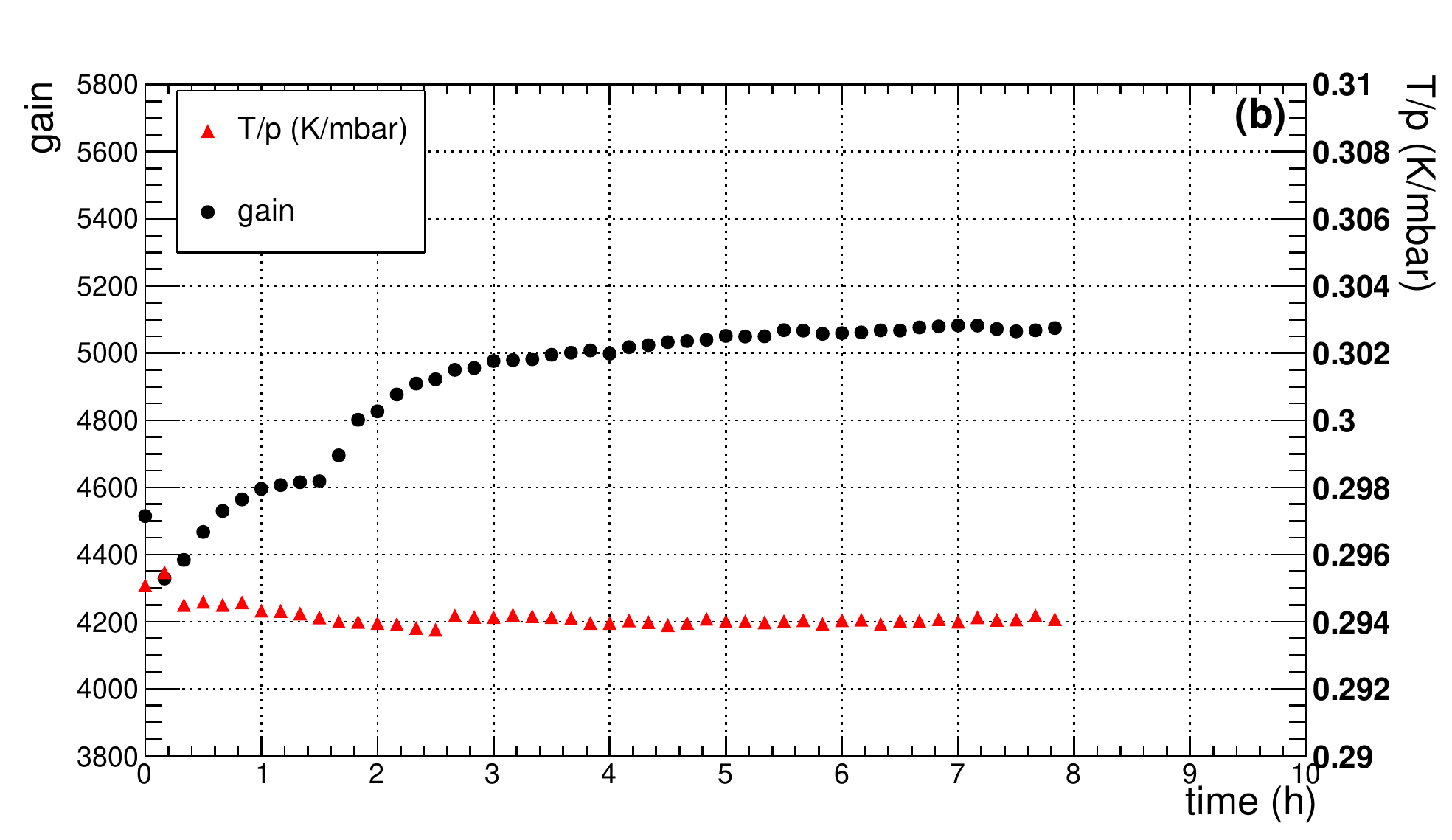}
		
		\includegraphics[scale=0.4]{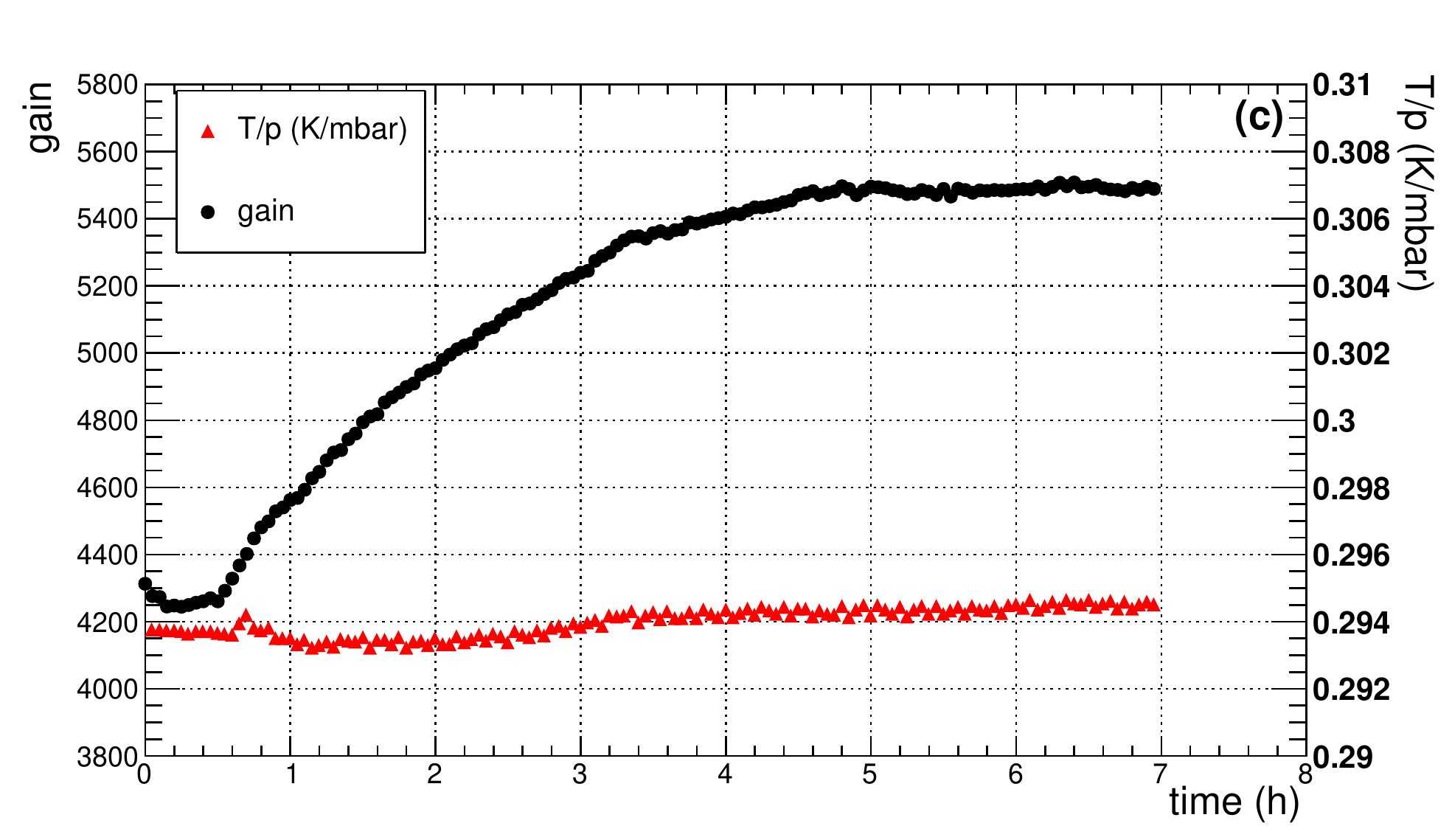}
		\caption{\label{fig:gain_tp_time}Variation of gain and T/p as a function of time. The top~(a), middle~(b) and bottom~(c) plots are for 1~kHz, 10~kHz and 90~kHz X-rays irradiation rates falling on 13$~mm^2$~(0.08~kHz/mm$^{2}$), 50$~mm^2$~(0.2~kHz/mm$^{2}$) and 28$~mm^2$~(3.2~kHz/mm$^{2}$) area of the GEM chamber respectively. All the measurement are caried out at a HV of -~4.2~kV and three different positions on the active area of the chamber.}
	\end{figure}
	%%%%%%%%%%%%%%%%%%%%%%%%%%%%%%%%%%%%%%%%%%%%%%%%%    
	
	Since it is well known that the gain of any gaseous detector depends on temperature and pressure~\cite{tp_gas_detector}, that is why the variation in T/p is plotted along with the gain as a function of time. The HV is kept OFF for 180 minutes, 60 minutes, and 8 minutes before the measurement is started with 1~kHz, 10~kHz, and 90~kHz X-ray rates respectively. In all three cases, the data taking starts immediately after the HV is switched ON and the source is placed on the active area of the detector. The energy spectra are stored at an interval of 10 minutes for 1~kHz and 10~kHz rates and 3 minutes for 90~kHz X-ray rates respectively. The same Fe$^{55}$ source is used to irradiate the chamber as well as to obtain the spectra. From Fig.~\ref{fig:gain_tp_time}, it is evident that the gain decreases for the first few minutes and then increases for a few hours of operation and reaches a saturation asymptotically. The decrease in the initial gain may be due to the loss of the primary electrons(/ions) which are stuck on the polarised dielectric~(Kapton) surface. Since the polarisation of the dielectric medium itself takes some finite time that is why whenever the HV and irradiation started simultaneously, an initial decrease in gain is observed. Afterward, the gain increases sharply for the first few hours due to the lensing effect created by the accumulated charges on the wall of the Kapton foil because this effect increases the electric field strength inside the GEM hole. The absolute gain values after saturation are not the same for all the cases due to the different source positions. The variation in gain over the active area of the particular chamber is reported earlier~\cite{s. chatterjee} and it was found that there was a variation of $\sim$10\%~(RMS). For all the measurements, the gain shows a saturation followed by an initial increase. The gain is normalised further to eliminate the T/p dependence on the gain. For the T/p normalisation, data obtained after $\sim$360 minutes of operation~(i.e. the saturated gain value) is used where only the T/p effect is dominant on the gain variation.  The method of normalisation is discussed in Ref~\cite{s. roy}. The normalised gain is fitted with an exponential function of the form~\cite{P. hauer}
	\begin{equation}
	\label{eqn}
	G = p_0(1-p_1e^{(-t/p_2)})\tag{1}
	\end{equation}
	where $G$ is the normalised gain, $p_0$ \& $p_1$ are the constants, $t$ is the measurement time in hours, and $p_2$ is the time constant of the charging-up effect, taking analogy from the charging up mechanism of any RC network~\cite{V. Tikhonov}.
	
	%%%%%%%%%%%%%%%%%%%%%%%%%%%%%%%%%%%%%%%%%%%%%%%%%    
	\begin{figure}[htbp]
		\centering
		\includegraphics[scale=0.4]{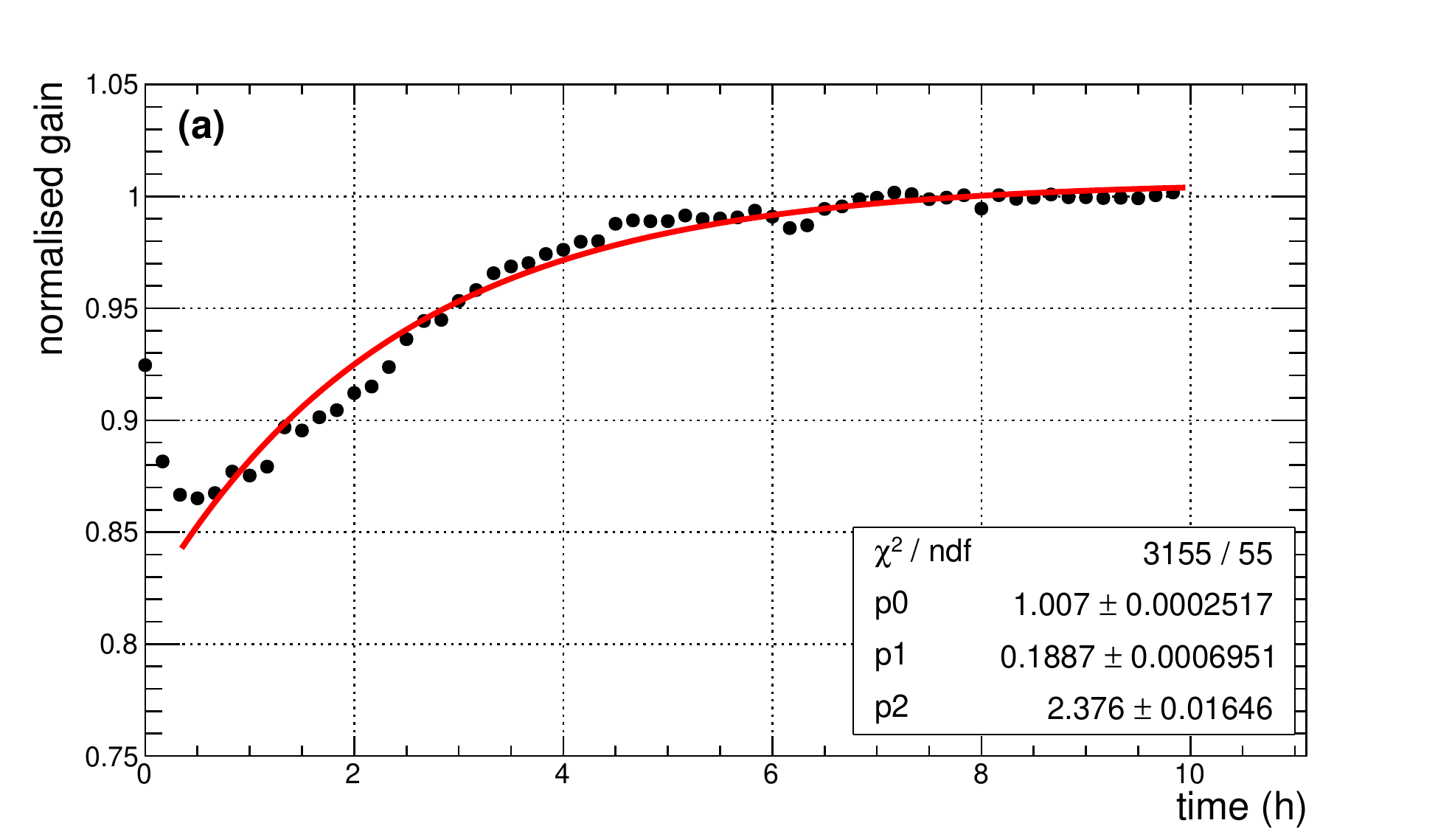}
		\includegraphics[scale=0.4]{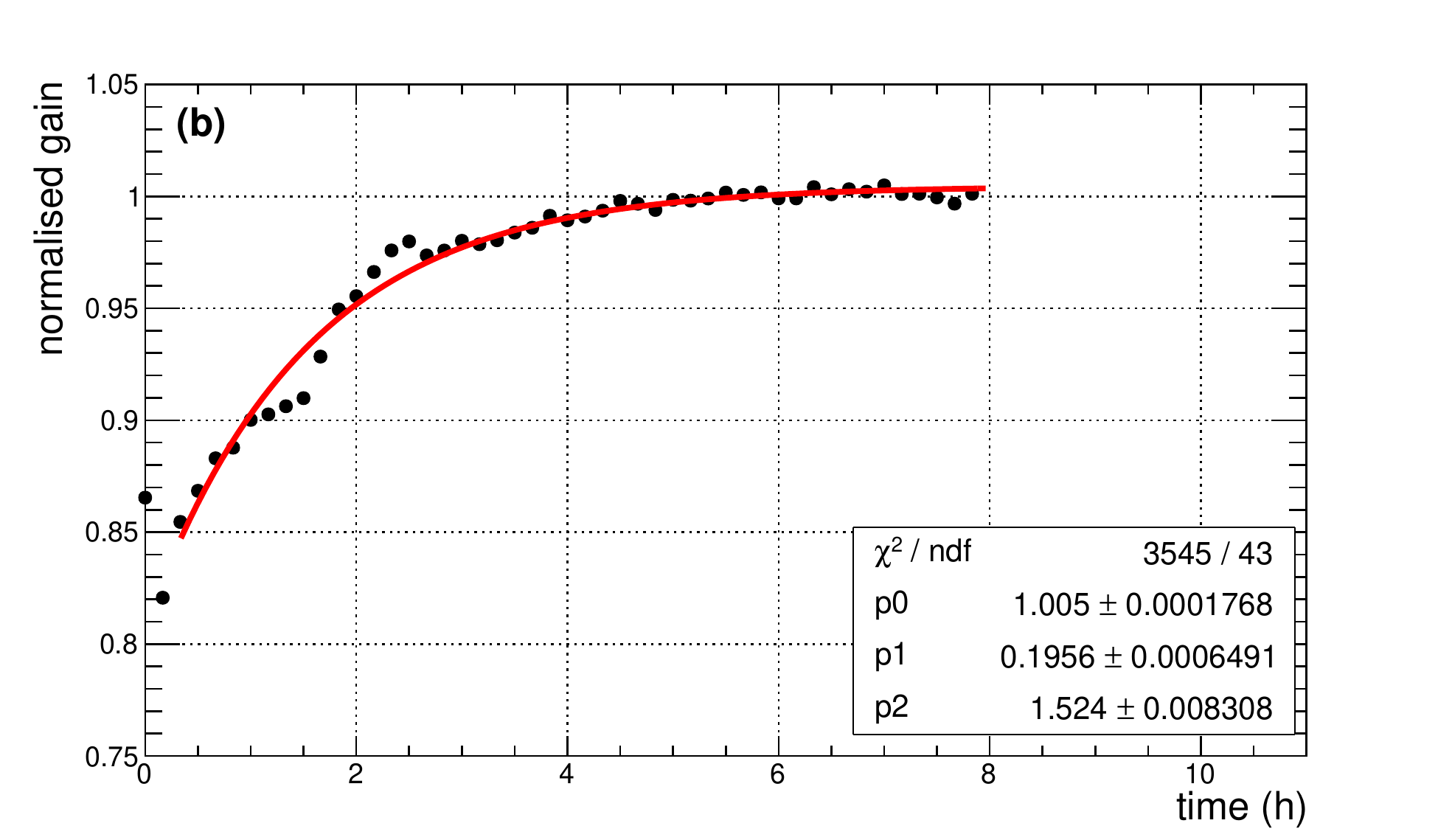}
		%\hspace*{3.7cm}
		\includegraphics[scale=0.4]{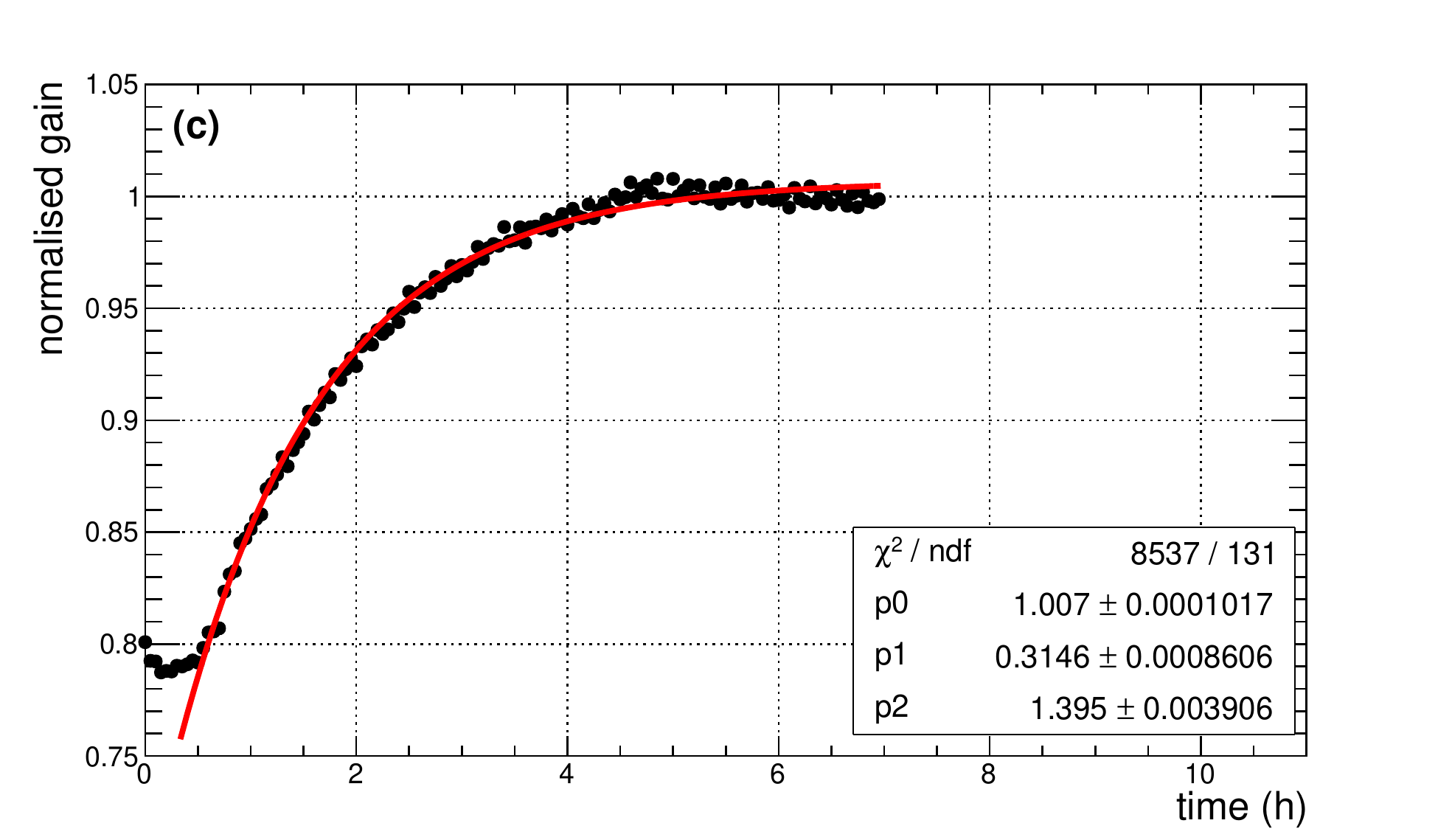}
		%\vspace{-0.4cm} 
		\caption{\label{fig:normalised_gain}Variation of the normalised gain as a function of time. The top~(a), middle~(b) and bottom~(c) plots are for 1~kHz, 10~kHz and 90~kHz X-rays irradiation rates falling on 13$~mm^2$~(0.08~kHz/mm$^{2}$), 50$~mm^2$~(0.2~kHz/mm$^{2}$), 28$~mm^2$~(3.2~kHz/mm$^{2}$) area of the GEM chamber respectively.}
	\end{figure}
	%%%%%%%%%%%%%%%%%%%%%%%%%%%%%%%%%%%%%%%%%%%%%%%%%
	%%%%%%%%%%%%%%%%%%%%%%%%%%%%%%%%%%%%%%%%%%%%%
	\begin{figure}[htbp]
		\centering

		\includegraphics[scale=0.40]{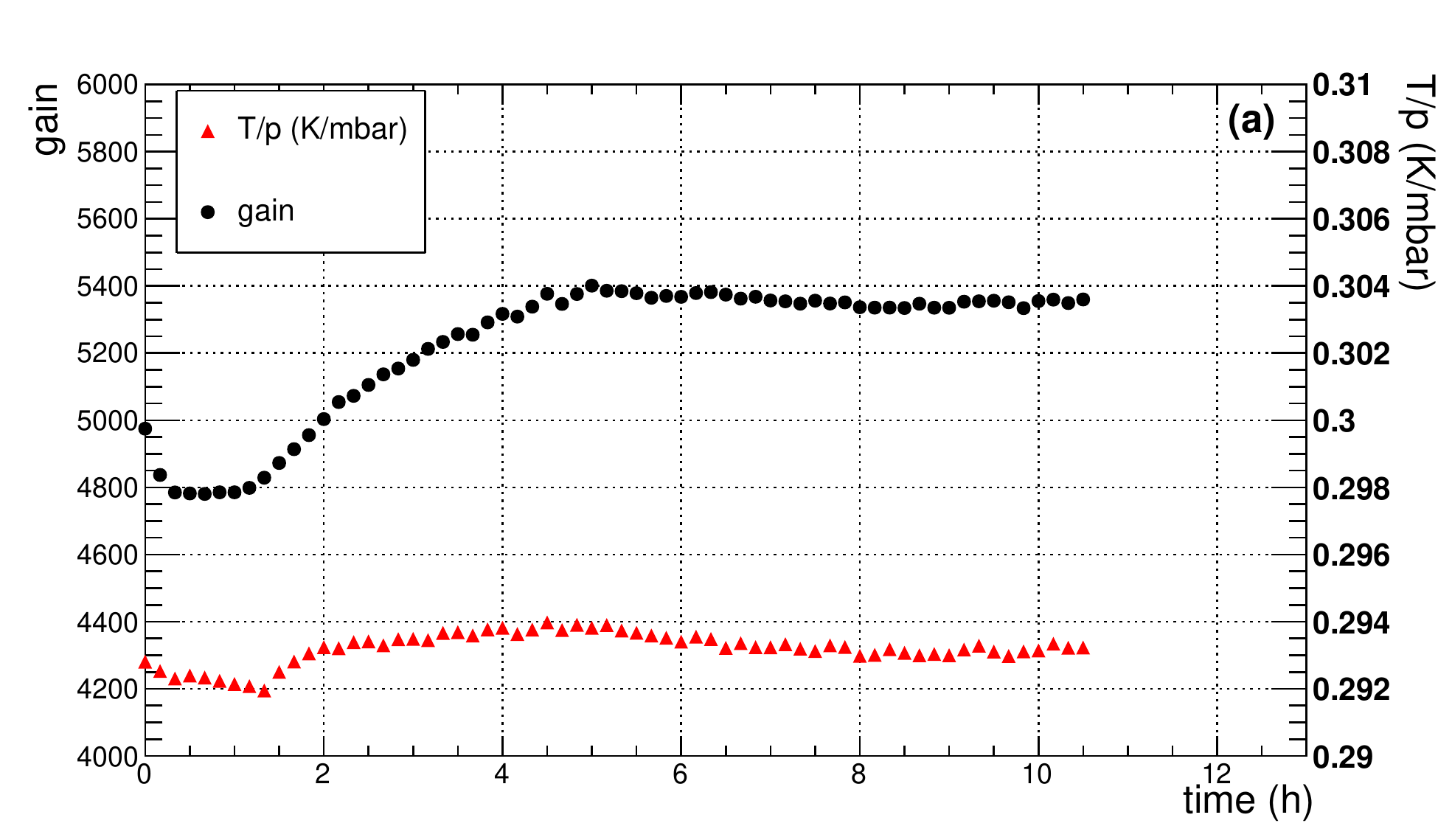}
		\includegraphics[scale=0.40]{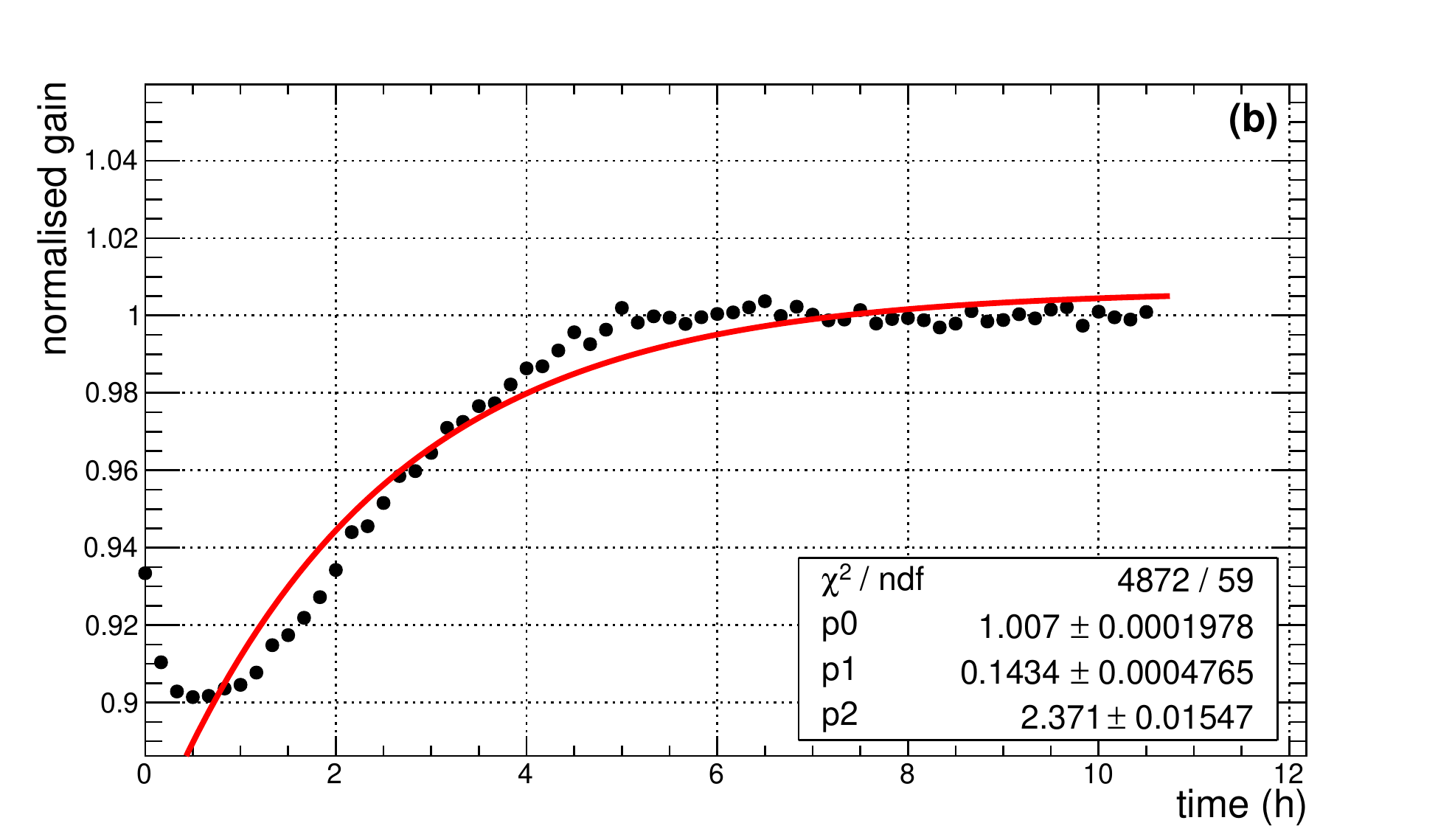}
		\caption{\label{fig:example}Variation of gain, T/p~(a) and normalised gain~(b) as a function of time for 1~kHz X-rays irradiating 13$~mm^2$~(0.08~kHz/mm$^{2}$) area of the GEM chamber. The measurement has been carried out at an HV of -~4.2~kV. The HV was kept OFF for $\sim$~60~minutes before taking the first measurement with the Fe$^{55}$ X-ray source.}
	\end{figure}
	%%%%%%%%%%%%%%%%%%%%%%%%%%%%%%%%%%%%%%%%%%%%%

	%%%%%%%%%%%%%%%%%%%%%%%%%%%%%%%%%%%%%%%%%%%%%%%%%
	\begin{figure}[htbp]
		\centering
		
		\includegraphics[scale=0.40]{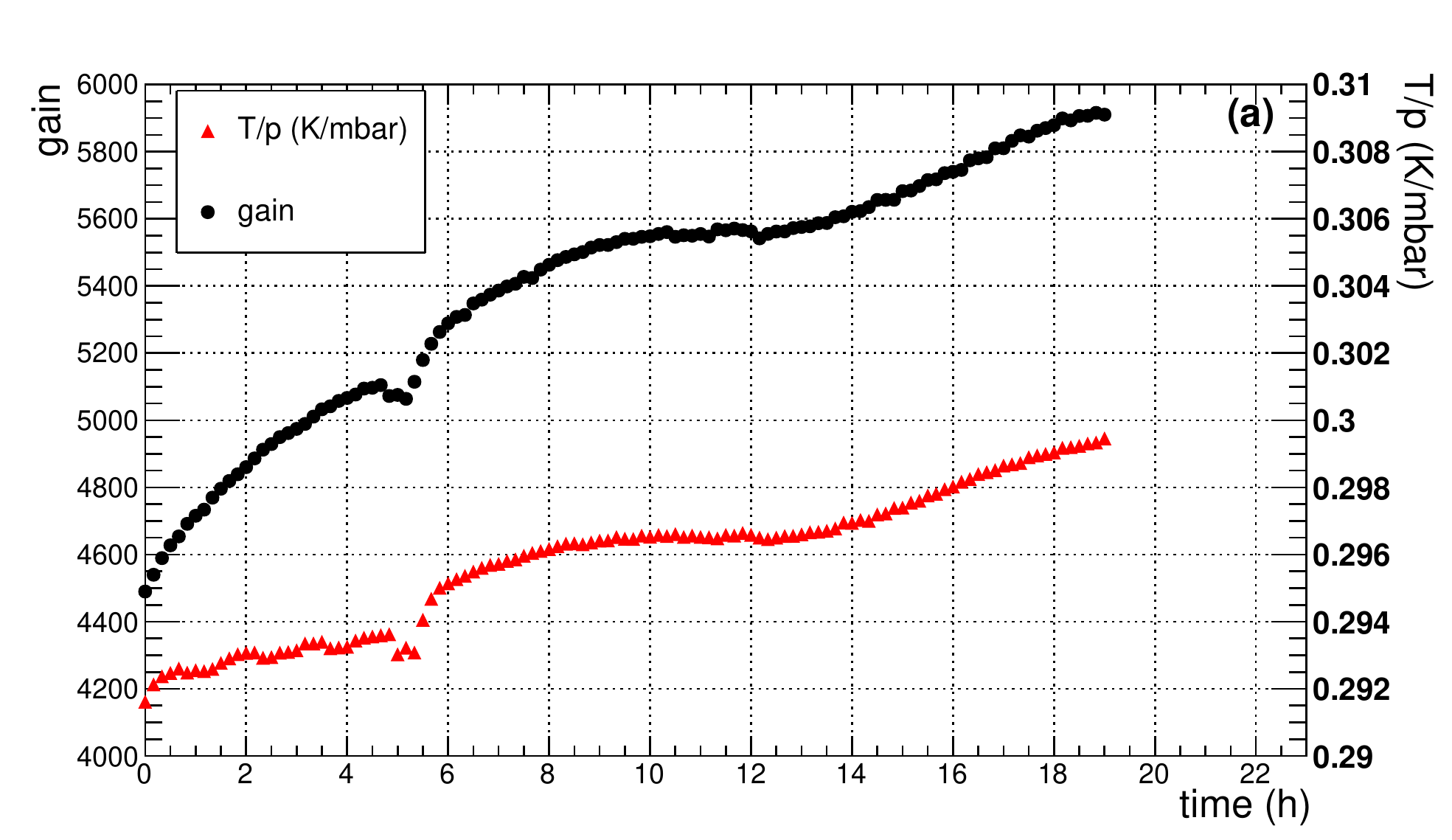}
		\includegraphics[scale=0.40]{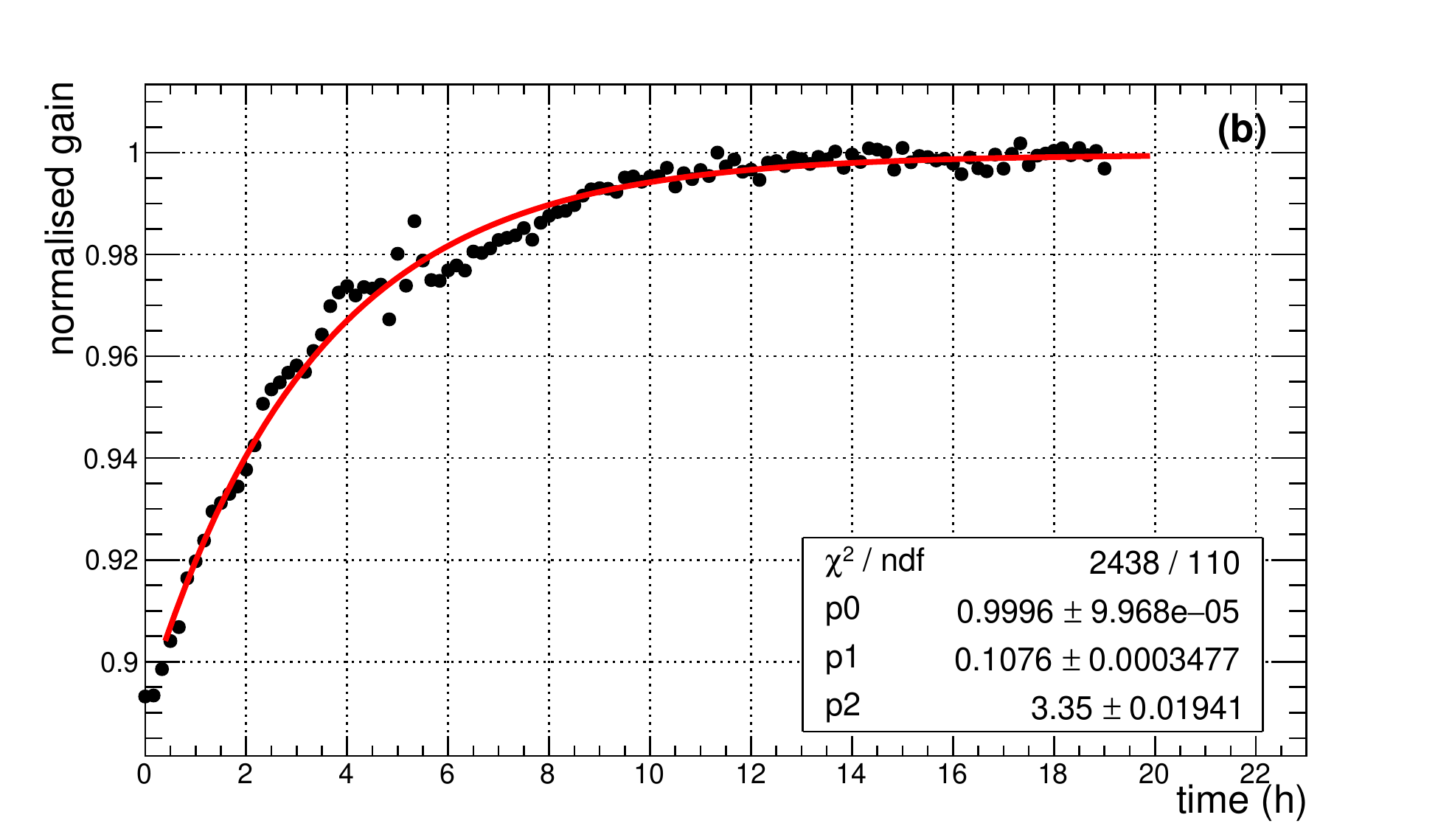}
		\caption{\label{fig:polarisation}Variation of gain, T/p~(a) and normalised gain~(b) as a function of time for 1~kHz X-rays irradiating 13$~mm^2$~(0.08~kHz/mm$^{2}$) area of the GEM chamber. The measurement is carried out at HV of - 4.1~kV. The HV is kept ON for 24 hours before taking the first measurement with the Fe$^{55}$ X-ray source.}
	\end{figure}
	%%%%%%%%%%%%%%%%%%%%%%%%%%%%%%%%%%%%%%%%%%%%%%%%%
	The fitted normalised gain is shown in Fig.~\ref{fig:normalised_gain} for the 1~kHz~(top), 10~kHz~(middle) and 90~kHz~(bottom) X-ray irradiation rates. For the fitting, the first $\sim$20 minutes are excluded because that includes both the effect of dielectric polarisation and charging up. After that, the charging up effect is dominant and is fitted with equation~\ref{eqn} to get an idea about the time constant of the charging up effect. In Fig.~\ref{fig:gain_tp_time}~(b), a small change is visible in the trend of increasing gain from 1-2 hours along the time axis and that is due to the two opposite effects namely charging-up and T/p variation. The charging-up process will tend to increase the gain and decrease in T/p will tend to reduce the gain. As a result of these two competing processes, the slope of the curve changes, and that is also reflected in Fig.~\ref{fig:normalised_gain}~(b). This competing effect between the charging-up and T/p is distinct in Fig.~\ref{fig:example} where 1~kHz X-rays irradiation 13~$mm^2$ area of the GEM chamber. For this measurement, the high voltage is switched ON, the source is placed on the detector, and data-taking is started. Before that, the HV is kept OFF for $\sim$ 60 minutes. The first three points in Fig.~\ref{fig:example}~(a) show a decreasing trend in the gain which is a combined effect of T/p and dielectric polarisation. After that, though the T/p value shows a decreasing trend, there is no visible decrease in the gain. That is due to the two competing processes, the effect of the decreasing T/p and charging-up on the gain is anti-correlated. Then after $\sim$1.5~hr, the gain increases because of the charging-up and T/p variation. The corresponding normalised gain variation is shown in Fig.~\ref{fig:example}~(b).

	To identify whether the decrease in the gain at the first few minutes is due to dielectric polarisation or not, a different measurement is performed by keeping the HV ON for $\sim$24 hours before the first measurement. The HV is kept at -~4.1~kV which corresponds to $\Delta V \sim$~382~V across each GEM foil and drift field, transfer field, and induction field of ~2.3~kV/cm,~3.4~kV/cm and~3.4~kV/cm respectively. The chamber is irradiated with 1~kHz X-ray falling on 13~mm$^2$ area of the chamber. Once the source is placed, the measurement is started immediately. The variation of gain, T/p, and normalised gain is shown as a function of time in Fig.~\ref{fig:polarisation}~(a) and~(b) respectively. The data is stored at an interval of 10~minutes. It is evident from the plot that there is no decrease in gain is observed at the beginning.  
	
	Since the charging-up process is due to the accumulation of the charges on the GEM holes therefore the charging-up process depends on the flux of incident radiation. More the flux of the incident particle faster will be the charging-up effect and the same behavior also appears from this study. From Fig.~\ref{fig:normalised_gain}, for 1~kHz, 10~kHz, and 90~kHz operations, the time constant of the charging-up effect is found to be 2.376~$\underline{+}$~0.02 hours, 1.524~$\underline{+}$~0.008 and 1.395~$\underline{+}$~0.004 hours respectively. The time constant of charging-up effect obtained from Fig.~\ref{fig:example}~(b), agrees well with~\ref{fig:normalised_gain}~(a). From Fig.~\ref{fig:polarisation}~(b), the time constant of the charging-up effect is found to be 3.294~$\underline{+}$~0.018 for 1~kHz X-ray. The time constant of charging-up effect obtained from~\ref{fig:polarisation}~(b) can not be compared with Fig.~\ref{fig:normalised_gain}~(a) and Fig.~\ref{fig:example}~(b)  because the HV is different. The residual voltage dependence on the charging-up effect is also seen in Ref~\cite{P. hauer}.
	
	\section{Summary and Outlook}
	The charging-up effect of a double mask triple GEM prototype is studied using different irradiation rates from a Fe$^{55}$ X-ray source. The chamber is operated with Ar/CO$_2$ gas mixture in a 70/30 volume ratio. The HV is kept OFF for a few minutes to several hours before starting the respective measurements. The data is stored just after the HV is ON and the source is placed on the chamber to see the effect of dielectric polarisation on the gain of the chamber. It is observed that the gain initially decreases and then increases to reach a saturation value. To ensure the decrease in the gain of the chamber during the first few minutes is due to dielectric polarisation, a different set of measurements is performed where the HV is kept ON for $\sim$24 hours before the data taking. No initial decrease in gain is observed in that case as shown in Fig.~\ref{fig:polarisation} because the dielectric (i.e. Kapton) is already polarised due to the application of HV beforehand. With different particle fluxes, the time constant of the charging up effect is investigated. It is found that the charging-up time decreases with increasing particle flux. Though the time constant value is decreasing with increasing particle flux, the exact scaling of the time constant with particle flux is not possible because we are observing an overall effect due to the three GEM foils and it is very difficult to disentangle the effects of each GEM foil on the final results. Also since the charging up time depends on the GEM hole geometry, properties of the Kapton foil, charge density in the GEM holes, etc, therefore we can only conclude that the time constant of the charging up effect decreases with increasing particle flux. The two competing effects of T/p variation and charging-up on the gain the chamber is studied by recording the 5.9~keV Fe$^{55}$ X-ray spectra and as expected it is coming to be anti-correlated as shown in Fig.~\ref{fig:example}. The dependence of the charging-up process on the gain of the detector, electric field strengths of different layers, used gas mixture, and also on the different kinds of GEM foils is under investigation.

\section{Acknowledgments}
	
	The authors would like to thank the RD51 collaboration for the support in building and initial testing of the chamber in the RD51 laboratory at CERN. We would like to thank Dr. A. Sharma, Dr. L. Ropelewski, Dr. E. Oliveri and Dr. Chilo Garabatos of CERN and Dr.~C.~J.~Schmidt and Mr.~J{\"o}rg~Hehner of GSI Detector Laboratory and  Prof. Sanjay K. Ghosh, Prof. Sibaji Raha, Prof. Rajarshi Ray and Dr. Sidharth K. Prasad of Bose Institute for valuable discussions and suggestions in the course of the study. This work is partially supported by the research grant SR/MF/PS-01/2014-BI from DST, Govt. of India, and the research grant of CBM-MuCh project from BI-IFCC, DST, Govt. of India. S. Chatterjee acknowledges his Institutional Fellowship research grant of Bose Institute. S. Biswas acknowledges the support of DST-SERB Ramanujan Fellowship (D.O. No. SR/S2/RJN-02/2012) and Intramural Research Grant provided by Bose Institute.

\end{document}